\documentclass[a4paper]{jpconf}

\usepackage{tikz}

\usepackage[pass]{geometry} 
\usepackage{color, braket,bm,subfig, mathrsfs,mathtools, wasysym,amsmath, amsfonts,amssymb,graphicx, slashed, cite}
  \DeclareMathAlphabet{\mathpzc}{OT1}{pzc}{m}{it}

\makeatletter

\def\ps@headings{%
     \def\@oddfoot{\hfil\thepage\hfil}
     \def\@evenfoot{\hfil\thepage\hfil}
     \let\@oddhead\@empty
     \let\@evenhead\@empty
      \let\@mkboth\markboth
      \let\sectionmark\@gobble
      \let\subsectionmark\@gobble}

\makeatother

\begin{document}

%%%%%%%%%%%%%%%%%% remove for submission %%%%%%%%%%%%%%%%

\begin{flushright}
\begin{footnotesize}
 MAN/HEP/2015/04\\
% March 2015
\end{footnotesize} 
\end{flushright}

%%%%%%%%%%%%%%%%%%%%%%%%%%%%%%%%%%%%%%%%%%%%%%%%%%%%%%%%%

\title{Natural Standard Model Alignment in the Two Higgs Doublet Model}

\author[a]{\underline{P.~S.~Bhupal Dev} and Apostolos Pilaftsis}

\address{Consortium for Fundamental Physics,
  School of Physics and Astronomy, University of Manchester, Manchester M13 9PL, United Kingdom.}
%%%%%%%%%%%%%%%%%%%%%%%%%%%%%%%%%%%%%%%%%%%%%%%%%%%%%%%%%%%%%%%%%%%%%%%%%%%
\begin{abstract}
The current LHC Higgs data provide strong constraints on possible deviations of the couplings of the observed 125 GeV Higgs boson from the Standard Model (SM) expectations. Therefore, it now becomes compelling that any extended Higgs sector must comply with the so-called SM {\em alignment limit}. In the context of the Two Higgs Doublet Model (2HDM), this alignment is often associated with either decoupling of the heavy Higgs sector or accidental cancellations in the 2HDM potential. Here we present a new solution realizing {\em natural} alignment based on symmetries, without decoupling or fine-tuning. In particular, we show that in 2HDMs where both Higgs doublets acquire  vacuum expectation values, there exist only {\em three} different symmetry realizations leading to natural alignment. We discuss some phenomenological implications of the Maximally-Symmetric 2HDM based on SO(5) symmetry group and analyze new collider signals for the heavy Higgs sector, involving third-generation quarks, which can be a useful observational tool  during the Run-II phase of the LHC.
\end{abstract}
%%%%%%%%%%%%%%%%%%%%%%%%%%%%%%%%%%%%%%%%%%%%%%%%%%%%%%%%%%%%%%%%%%%%%%%%%%%%%
\section{Introduction}\label{sec:1}
The discovery of a Higgs boson with mass around 125 GeV~\cite{Aad:2012tfa} is the main highlight of the Run-I phase of the LHC, as it provides the first experimental evidence of the Higgs mechanism~\cite{Higgs:1964pj}. Although the
measured  couplings of  the  discovered Higgs  boson show  remarkable
compatibility  with  those   predicted   by  the   Standard Model (SM)~\cite{coup},  
the current  experimental  data still  leave open  the
possibility of an extended Higgs sector. 
In  fact,  several  well-motivated  new-physics scenarios  necessarily come with an
enlarged Higgs  sector, such as  supersymmetry~\cite{Haber:1984rc} and axion models~\cite{Kim:1986ax}, in order
to address a number  of theoretical and cosmological issues, including
the gauge  hierarchy problem,  the origin of  the Dark Matter (DM) and matter-antimatter 
asymmetry in  our Universe. Here we consider the simplest extension of
the standard Higgs mechanism, namely the Two Higgs Doublet Model~(2HDM)~\cite{review}, where the SM Higgs doublet is supplemented by another isodoublet with hypercharge $Y=1$. This model can provide new sources of spontaneous~\cite{Lee:1973iz} or explicit~\cite{Georgi:1978xz} CP violation, viable DM candidates~\cite{Silveira:1985rk} and a strong first order phase transition for electroweak baryogenesis~\cite{Kuzmin:1985mm}.

In the doublet field space $\Phi_{1,2}$, where $\Phi_i = (\phi_i^+, \phi_i^0)^{\sf T}$, the  general 2HDM  potential reads 
\begin{align}
  \label{pot}
V \ = \ & -\mu_1^2(\Phi_1^\dag \Phi_1) -\mu_2^2 (\Phi_2^\dag \Phi_2) 
-\left[m_{12}^2 (\Phi_1^\dag \Phi_2)+{\rm H.c.}\right] \nonumber \\ &
+\lambda_1(\Phi_1^\dag \Phi_1)^2+\lambda_2(\Phi_2^\dag \Phi_2)^2 
 +\lambda_3(\Phi_1^\dag \Phi_1)(\Phi_2^\dag \Phi_2) 
 +\lambda_4(\Phi_1^\dag \Phi_2)(\Phi_2^\dag \Phi_1) \nonumber \\
& 
+\left[\frac{\lambda_5}{2}(\Phi_1^\dag \Phi_2)^2%\right.\\ 
%&&\left. 
+\lambda_6(\Phi_1^\dag \Phi_1)(\Phi_1^\dag \Phi_2) 
+\lambda_7(\Phi_1^\dag \Phi_2)(\Phi_2^\dag \Phi_2)+{\rm H.c.}\right],
\end{align}
which contains  four real mass  parameters $\mu_{1,2}^2$, Re$(m^2_{12})$,
Im$(m^2_{12})$,  and ten  real  quartic couplings  $\lambda_{1,2,3,4}$,
Re($\lambda_{5,6,7})$,  and Im($\lambda_{5,6,7}$). Thus,
the   vacuum   structure   of   the   general  2HDM   can   be   quite
rich~\cite{Battye:2011jj}, as compared to the SM. 
% and in  principle,  can allow  for a  wide
%range  of  parameter space  still  compatible  with  the existing  LHC
%constraints.    

The quark-sector Yukawa Lagrangian in the general 2HDM is given by 
\begin{align}
-{\cal L}^q_Y \  = \  \bar{Q}_L(h_1^u \Phi_1+ h_2^u\Phi_2)u_R \: +
\: \bar{Q}_L(h_1^d \widetilde{\Phi}_1+ h_2^d \widetilde{\Phi}_2)d_R \; ,
\label{yuk}
\end{align}
where $\widetilde{\Phi}_i={\rm i}\sigma_2\Phi_i^*$ are the isospin conjugates of $\Phi_i$, $Q_L=(u_L,d_L)^{\sf T}$ is the $SU(2)_L$ quark doublet and $u_R,d_R$ are 
right-handed quark singlets. Due to the Yukawa interactions in~\eqref{yuk},  
the neutral scalar bosons often induce unacceptably large 
flavor-changing neutral current (FCNC) processes at the tree level. This is usually avoided by 
imposing a discrete $Z_2$ symmetry~\cite{gw1} under which 
\begin{align}
\Phi_1 \to -\Phi_1,\quad \Phi_2\to \Phi_2, \quad u_{Ra}\to u_{Ra},\quad 
d_{Ra}\to d_{Ra}~~{\rm or}~~ d_{Ra}\to -d_{Ra} \; ,
\label{discrete}
\end{align}
($a=1,2,3$ being the quark family index) so that only $\Phi_2$ gives mass to up-quarks, and 
only $\Phi_1$ or only $\Phi_2$ gives mass to down-quarks. In this case, the scalar boson 
couplings to quarks are proportional to the quark mass matrix, as in the SM, and therefore, there is no tree-level FCNC process. The $Z_2$ symmetry~\eqref{discrete} is satisfied  by four  discrete  
choices of  tree-level Yukawa  couplings between the Higgs doublets and  SM fermions, which are known as Type I, II, X (lepton-specific) and Y (flipped) 2HDMs~\cite{review}. In Type II, X and  Y 2HDM, both Higgs doublets $\Phi_{1,2}$ acquire vacuum expectation values (VEVs) $v_{1,2}$, whereas in Type I 2HDM, one of the Higgs doublets ($\Phi_1$) does not couple to the SM fermions and need not acquire a VEV~\cite{Haber:1978jt}. Global fits to the current LHC Higgs data~\cite{fit1,fit2, Eberhardt:2013uba} suggest that all four types of discrete 2HDM are constrained to
lie close to the so-called SM {\it alignment limit}, where the mass
eigenbasis of the  CP-even scalar sector aligns with  the SM gauge
eigenbasis.   
%Specifically,  in  the  Type-II  (MSSM-type)  2HDM,  the
%coupling of the  SM-like Higgs to vector bosons  is constrained to lie
%within  10\%   of  the   SM  value  at   95\%  CL~\cite{Eberhardt:2013uba}.

Naively,  the SM alignment is often associated with  the decoupling limit,  
in which all  the non-standard Higgs bosons are assumed to be much heavier 
than the electroweak scale
so  that the  lightest CP-even  scalar behaves just like the  SM Higgs
boson.   This  SM  alignment  limit  can  also  be  achieved,  without
decoupling~\cite{Chankowski:2000an, Gunion:2002zf,  Ginzburg:2004vp, Delgado:2013zfa, Carena:2013ooa, Bhattacharyya:2014oka}. 
However, for  small $\tan\beta$ values, this is  usually attributed to
accidental cancellations in the 2HDM potential~\cite{Carena:2013ooa}. 
Here we present a symmetry argument to naturally
justify the alignment limit~\cite{Dev:2014yca},  independently of the
kinematic parameters  of the  theory, such as  the heavy  Higgs masses 
and the ratio of the VEVs ($v_2/v_1$).  In the 2HDMs where both Higgs doublets acquire VEVs, we show that there exist only 
{\em three} possible symmetry realizations of the scalar potential having natural alignment.   
We explicitly analyze the simplest case, namely the maximally symmetric 2HDM (MS-2HDM) 
with SO(5) symmetry. We show that the renormalization group (RG) effects
due to the hypercharge gauge coupling $g'$ and third-generation Yukawa
couplings,   as  well  as   soft-breaking  mass   parameters,  induce  
relevant deviations   from   the  SO(5)   limit, which lead to distinct predictions for
the Higgs  spectrum of the MS-2HDM.  In particular, the heavy Higgs sector is predicted to be {\em quasi-degenerate}, which is a distinct feature of the SO(5) limit, apart from being {\em gaugephobic}, which is a generic feature in the alignment limit. Moreover, the current experimental constraints force the heavy Higgs sector to lie above the top-quark threshold in the MS-2HDM. Thus, the 
dominant collider signal for this sector involves final states with third-generation quarks. 
We study some of these collider signals for the upcoming run of the LHC.

The plan of this proceedings is as follows: In Section~\ref{sec:2}, we present the natural alignment condition for a generic 2HDM scalar potential. In Section~\ref{sec:3}, we list the symmetry classifications of the 2HDM potential and identify the symmetries leading to a natural alignment. In Section~\ref{sec:4}, we analyze the MS-2HDM in presence of custodial symmetry and soft breaking effects. In Section~\ref{sec:5}, we discuss some collider phenomenology of the heavy Higgs sector in the alignment limit, with particular emphasis on the heavy Higgs sector beyond the top-quark threshold. Our conclusions are given in Section~\ref{sec:6}.

%%%%%%%%%%%%%%%%%%%%%%%%%%%%%%%%%%%%%%%%%%%%%%%%%%%%%%
\section{Natural Alignment Condition}\label{sec:2}
%%%%%%%%%%%%%%%%%%%%%%%%%%%%%%%%%%%%%%%%%%%%%%%%%%%%%%
For simplicity, we will consider the 2HDM potential~\eqref{pot} with CP-conserving vacua; the results derived in this section can be easily generalized to the CP-violating 2HDM potential. We start with the linear decomposition of the two Higgs doublets in terms of eight real scalar fields:
\begin{eqnarray}
\Phi_j \ = \ \left(\begin{array}{c} \phi_j^+ \\ \frac{1}{\sqrt
    2}(v_j+\phi_j+ia_j) \end{array} \right)\; , 
\label{expand-phi}
\end{eqnarray}
where $v=\sqrt{v_1^2+v_2^2}=246.2$~GeV  is the SM electroweak VEV.  After symmetry breaking, the three Goldstone   modes  ($G^\pm,G^0$)  become  the   longitudinal
components  of the $W^\pm$ and  $Z$ bosons, and there remain five
physical scalar mass eigenstates: two CP-even ($h,H$), one
CP-odd ($a$)  and two charged  ($h^\pm$) scalars. The corresponding physical      mass     eigenvalues     are     given
by~\cite{Haber:1993an,Pilaftsis:1999qt}
\begin{subequations}
\begin{align}
M^2_{h^\pm} \ & = \ \frac{m_{12}^2}{s_\beta
  c_\beta}-\frac{v^2}{2}\left( \lambda_4+ \lambda_5\right)
 +\frac{v^2}{2s_\beta c_\beta}\left( \lambda_6
c_\beta^2+ \lambda_7 s_\beta^2\right), \label{mass0}\\
%\nonumber \\ 
M_a^2 \ &  = \  M^2_{h^\pm}+\frac{v^2}{2}\left( \lambda_4 - 
\lambda_5\right), \label{mass1} \\ %\nonumber\\ 
M_H^2 \ & = \ \frac{1}{2}\left[(A+B)-\sqrt{(A-B)^2+4C^2}\right], \label{mass2}\\ %\nonumber\\
M_h^2 \ & = \ \frac{1}{2}\left[(A+B)+\sqrt{(A-B)^2+4C^2}\right], \label{mass3}
\end{align} 
\end{subequations}
where we have used the short-hand notations $c_\beta\equiv \cos\beta$ and $s_\beta\equiv \sin\beta$ with $\tan\beta=v_2/v_1$, and    
\begin{subequations}
\begin{align}
A \ & = \ M_a^2s_\beta^2+v^2\left( 2\lambda_1 c_\beta^2+ 
\lambda_5 s^2_\beta+2 \lambda_6 s_\beta
  c_\beta\right), \\ %\nonumber \\ 
B \ & = \ M_a^2c_\beta^2+v^2\left( 2\lambda_2 s_\beta^2+ 
\lambda_5 c^2_\beta + 2\lambda_7 s_\beta c_\beta
  \right) , \\ 
C \ & = \ -M_a^2 s_\beta c_\beta + v^2\left( \lambda_{34}s_\beta c_\beta
  + \lambda_6 c^2_\beta + \lambda_7 s^2_\beta
  \right) . 
\end{align}
\end{subequations}
with $\lambda_{34}=\lambda_3+\lambda_4$. The mixing between the mass eigenstates in the
CP-odd  and charged  sectors is  governed by the angle $\beta$, whereas in the CP-even sector, it is governed by the angle $\alpha$, where 
$\tan{2\alpha} = 2C/(A-B)$. 

The SM Higgs field can be identified as the linear combination  
\begin{eqnarray}
H_{\rm SM} \ = \  \phi_1 \cos\beta + \phi_2 \sin \beta \ = \ 
H\cos(\beta-\alpha)+h\sin(\beta-\alpha) \; . 
\label{HSM}
\end{eqnarray}
From~(\ref{HSM}), we see that the  couplings of  $h$ and $H$  to the  gauge bosons
($V=W^\pm, Z$) with  respect to the SM Higgs  couplings $g_{H_{\rm SM}VV}$ will be 
\begin{eqnarray}
g_{hVV} \ = \ \sin{(\beta-\alpha)} \; , \qquad g_{HVV} \ = \ \cos{(\beta-\alpha)} \; . 
\label{coup1}
\end{eqnarray}
Thus, the {\em SM alignment limit} is defined as the limit $\alpha\to \beta$ (or $\alpha\to \beta-\pi/2$) when $H$ ($h$) couples to the vector bosons exactly like in the SM, whereas $h$ ($H$) becomes gaugephobic.  For notational clarity, we will take the alignment limit to be $\alpha\to \beta$ in the following.

To derive the alignment condition, we rewrite the CP-even scalar mass matrix as
%\begin{widetext}
\begin{eqnarray}
  \label{align2}
M^2_{S} \  = \ \left(\begin{array}{cc} A& C \\ C & B \end{array}\right) 
\  = \ \left(\begin{array}{cc}
c_\beta & -s_\beta \\ 
s_\beta & c_\beta 
\end{array}\right) \left(\begin{array}{cc}
\widehat{A} & \widehat{C} \\
\widehat{C} & \widehat{B}
\end{array}\right)
\left(\begin{array}{cc}
c_\beta & s_\beta \\ 
-s_\beta & c_\beta 
\end{array}\right)\; , 
\end{eqnarray}
where 
\begin{subequations}
\begin{align}
\widehat{A} &  =  2v^2 \Big[ c_\beta^4 \lambda_1 
+ s_\beta^2 c_\beta^2 \lambda_{345}  
+ s_\beta^4 \lambda_2\:  +\: 2 s_\beta c_\beta \Big( c^2_\beta \lambda_6 +
s^2_\beta \lambda_7\Big)\Big]\; , \label{ahat} \\
\widehat{B} &  =   M_a^2\: +\: \lambda_5 v^2\: +\: 2 v^2 
\Big[ s^2_\beta c^2_\beta
  \Big(\lambda_1+\lambda_2-\lambda_{345}\Big)\:
-\: s_\beta c_\beta \Big(c^2_\beta - s^2_\beta\Big) \Big(\lambda_6 
- \lambda_7\Big) \Big]\; , \label{bhat} \\
\widehat{C} &  =   
v^2 \Big[ s^3_\beta c_\beta \Big( 2\lambda_2-\lambda_{345}\Big) - 
c^3_\beta s_\beta \Big(2\lambda_1- \lambda_{345}\Big) + c^2_\beta
\Big( 1 - 4 s^2_\beta \Big) \lambda_6 
+ s^2_\beta \Big( 4 c^2_\beta - 1\Big) \lambda_7 \Big]. \label{chat}
\end{align}
\end{subequations}
Here we have used  the  short-hand  notation:  $\lambda_{345} \equiv  \lambda_3  +
\lambda_4    +    \lambda_5$.     
%Observe    that    $\widehat{M}^2_S$
%in~(\ref{align2}) is the respective  $2\times 2$ CP-even mass matrix
%written down in the so-called Higgs eigenbasis~\cite{Georgi, Dono, Lavoura, Botella:1994cs}.
Evidently, the SM alignment limit $\alpha \to \beta$ is obtained when 
$\widehat{C} =  0$ in~\eqref{align2}~\cite{Gunion:2002zf}. From (\ref{chat}), this yields the quartic equation
\begin{eqnarray}
\lambda_7 t_\beta^4 -  (2\lambda_2-\lambda_{345})t_\beta^3 + 3(\lambda_6-\lambda_7)t_\beta^2 + 
(2\lambda_1-\lambda_{345})t_\beta - \lambda_6 \ = \ 0 \; .
\label{align-gen}
\end{eqnarray}
For {\em natural} alignment, (\ref{align-gen}) should be satisfied for {\it any} value of $\tan\beta$, which requires the coefficients of the polynomial in $\tan\beta$ to vanish identically.  Imposing this restriction, we arrive at the {\em natural alignment condition}~\cite{Dev:2014yca}
\begin{eqnarray}
2\lambda_1 \ = \ 2\lambda_2 \ = \ \lambda_{345}\;, \qquad \lambda_6 \ = \ \lambda_7 \ = \ 0\; . 
\label{alcond}
\end{eqnarray}
In particular, for $\lambda_6 = \lambda_7 = 0$ as in the $Z_2$-symmetric 2HDMs, (\ref{align-gen}) has a solution 
\begin{eqnarray}
  \label{tanb}
\tan^2\beta\ =\ \frac{2\lambda_1 - \lambda_{345}}{2\lambda_2 -
\lambda_{345}}\ >\ 0 \; ,
\end{eqnarray}
independent of $M_a$. After some algebra, the simple solution (\ref{tanb}) to our general alignment condition (\ref{align-gen}) can be shown to be equivalent to that derived in~\cite{Carena:2013ooa, Carena:2014nza}.

%%%%%%%%%%%%%%%%%%%%%%%%%%%%%%%%%%%%%%%%%%%%%%%%%%%%%%%%%%%%%%%%%%%%%%%%%%%%
\section{Symmetry Classifications of the 2HDM Potential}\label{sec:3}
%%%%%%%%%%%%%%%%%%%%%%%%%%%%%%%%%%%%%%%%%%%%%%%%%%%%%%%%%%%%%%%%%%%%%%%%%%%%
The general 2HDM potential~\eqref{pot} may exhibit three different classes of accidental symmetries. The first class of symmetries pertains to transformations of the Higgs doublets $\Phi_{1,2}$ only, but not their complex conjugates $\Phi_{1,2}^*$, and are known as the Higgs family (HF) symmetries~\cite{Ginzburg:2004vp, Ferreira:2009wh}.  The second class of symmetry transformations relates the fields $\Phi_{1,2}$ to their complex conjugates $\Phi_{1,2}^*$ and are generically termed as CP symmetries~\cite{Ferreira:2009wh}. The third class of symmetries utilize mixed HF and CP transformations that leave the $SU(2)_L$ gauge kinetic terms of $\Phi_{1,2}$ canonical~\cite{Battye:2011jj}.   
 
To identify all  accidental symmetries of the 2HDM potential,
it   is   convenient    to  work in the bilinear scalar field formalism~\cite{Maniatis:2006fs} by introducing  an  8-dimensional   complex multiplet~\cite{Battye:2011jj,Nishi:2011gc,Pilaftsis:2011ed}: 
\begin{equation}
  \label{Phi}
{\bf \Phi}\ \equiv\ \left(\begin{array}{c}  
\Phi_1  \\
\Phi_2  \\  
\widetilde{\Phi}_1  \\
\widetilde{\Phi}_2 \end{array}\right)\; ,
\end{equation}
where $\widetilde{\Phi}_{i} =  {\rm i}\sigma^2 \Phi^*_{i}$ (with $i=1,2$) and $\sigma^{2}$  is the second  Pauli matrix. 
%We
%should remark  that the complex  multiplet ${\bf \Phi}$  satisfies the
%Majorana property~\cite{Battye:2011jj}: ${\bf \Phi} = C {\bf \Phi}^*$,
%where  $C  =  \sigma^2  \otimes  \sigma^0  \otimes  \sigma^2$  is  the
%charge-conjugation matrix, with $\sigma^0 =
%{\bf 1}_{2\times  2}$ being the  identity matrix. 
In terms of the ${\bf \Phi}$-multiplet, the following   {\em   null}   6-dimensional   Lorentz   vector   can   be defined~\cite{Battye:2011jj, Pilaftsis:2011ed}:
\begin{equation}
  \label{RA}
R^A\ \equiv\ {\bf \Phi}^\dag \Sigma^A {\bf \Phi}\; ,
\end{equation}  
where  $A=0,1,...,5$  and  the  six $8\times  8$-dimensional  matrices
$\Sigma^A$ may be expressed in terms of the three Pauli matrices $\sigma^{1,2,3}$ and the identity matrix $\bm{1}_{2\times 2}\equiv \sigma^0$, as follows:
\begin{eqnarray}
&& \Sigma^{0,1,3} \ = \ \frac{1}{2}\sigma^0 \otimes \sigma^{0,1,3} \otimes
  \sigma^0, \quad   
\Sigma^2 \ = \ \frac{1}{2}\sigma^3 \otimes \sigma^2 \otimes \sigma^0, \nonumber\\
&& \Sigma^4 \ = \ -\frac{1}{2}\sigma^2 \otimes \sigma^2 \otimes \sigma^0, \quad
\Sigma^5 \ = \ -\frac{1}{2}\sigma^1 \otimes \sigma^2 \otimes \sigma^0. 
\end{eqnarray}
Note that the bilinear  field space spanned by the  6-vector $R^A$ realizes
an {\em orthochronous} ${\rm SO}(1,5)$ symmetry group.

In terms of  the null-vector $R^A$ defined in~(\ref{RA}), the  2HDM potential (\ref{pot}) 
takes on a simple quadratic form:
\begin{equation}
  \label{potR}
V\ =\ -\, \frac{1}{2}\, M_A\,R^A\: + \: \frac{1}{4}\, L_{AB}\, R^A R^B\; ,
\end{equation}
where $M_A$  and $L_{AB}$ are ${\rm SO}(1,5)$  constant `tensors' that
depend  on the  mass parameters  and quartic  couplings given in~\eqref{pot}   
 and    their   explicit    forms    may   be    found
in~\cite{Pilaftsis:2011ed,Maniatis:2007vn}.
Requiring   that    the   SU(2)$_L$   gauge-kinetic    term   of   the 
{\boldmath $\Phi$}-multiplet  remains canonical restricts  the allowed
set of  rotations from SO(1,5)  to SO(5),  where only  the  spatial components
$R^I$ (with $I=1,...,5$) transform and the zeroth component $R^0$
remains invariant. Consequently, in  the absence of  the hypercharge
gauge coupling and fermion Yukawa couplings, the maximal symmetry
group of the 2HDM is $G^R_{\rm  2HDM} = {\rm SO(5)}$. Including all its proper, improper 
and semi-simple subgroups of SO(5), all accidental symmetries for the 2HDM 
potential were classified in~\cite{Battye:2011jj, Pilaftsis:2011ed}, as shown in Table~\ref{tab1}. Here we have used a diagonally reduced basis~\cite{Gunion:2005ja}, where ${\rm Im}(\lambda_5)=0$ and $\lambda_6=\lambda_7$, thus reducing the number of independent quartic couplings to seven. Each of the symmetries listed in Table~\ref{tab1} leads to certain constraints on the mass and/or coupling parameters.   
\begin{table}[t!]
\begin{center}
\begin{tabular}{c|ccccccccc}\hline
symmetry & $\mu_1^2$ & $\mu_2^2$ & $m^2_{12}$ & $\lambda_1$ & $\lambda_2$ & $\lambda_3$ & $\lambda_4$ & ${\rm Re}(\lambda_5)$ & $\lambda_6=\lambda_7$ \\ \hline
$Z_2\times$ O(2) & - & - & Real & -& -& -& -& -& Real\\
$(Z_2)^2\times $SO(2) & -&  - & 0 & - & -&  - & - & - & 0 \\
$(Z_2)^3\times $O(2) & -&  $\mu_1^2$ & 0 & - &  $\lambda_1$ &  - & - & - & 0 \\
O(2) $\times $O(2) & -&  - & 0 & - & -&  - & - & 0 & 0 \\
$Z_2\times $ [O(2)]$^2$ & -& $\mu_1^2$ & 0 & - & $\lambda_1$ &  - & - & $2\lambda_1-\lambda_{34}$ & 0 \\
O(3)$\times $O(2) & -&  $\mu_1^2$ & 0 & - & $\lambda_1$ &  - & $2\lambda_1-\lambda_3$  & 0 & 0 \\
SO(3) & -&  - & Real & - & - &  - & -  & $\lambda_4$ & Real \\
$Z_2\times $O(3) & - &  $\mu_1^2$ & Real & - & $\lambda_1$ &  - & -  & $\lambda_4$ & Real \\
$(Z_2)^2\times $SO(3) & -&  $\mu_1^2$ & 0 & - & $\lambda_1$ &  - & -  & $\pm \lambda_4$ & 0 \\
O(2)$\times $O(3) & -&  $\mu_1^2$ & 0 & - & $\lambda_1$ &  $2\lambda_1$ & -  & 0 & 0 \\
SO(4) & -&  - & 0 & - & - &  - & 0  & 0 & 0 \\
$Z_2\times $O(4) & -&  $\mu_1^2$ & 0 & - & $\lambda_1$ &  - & 0  & 0 & 0 \\
SO(5) & -&  $\mu_1^2$ & 0 & - & $\lambda_1$ & $2\lambda_1$ & 0 & 0 & 0 \\ \hline
\end{tabular}
\end{center}
\caption{Relations between the parameters of the $U(1)_Y$-invariant 2HDM potential~\eqref{pot} for the 13 accidental symmetries~\cite{Pilaftsis:2011ed} in a diagonally reduced basis, where ${\rm Im}(\lambda_5)=0$ and $\lambda_6=\lambda_7$. A dash signifies the absence of a constraint for that parameter. Notice that all symmetries lead to a CP-conserving 2HDM potential.}\label{tab1}
\end{table}

From Table~\ref{tab1}, we observe that there are {\it only} three symmetries, namely (i) $Z_2\times [{\rm O(2)}]^2$, (ii) O(3)$\times$ O(2) and (iii) SO(5), which satisfy the natural alignment condition given by~\eqref{alcond}.\footnote{In Type-I 2HDM, there exists an additional possibility of realizing an exact Z$_2$ symmetry~\cite{Deshpande:1977rw} which leads to an exact alignment, i.e. in the context of the so-called inert 2HDM~\cite{Barbieri:2006dq}.} Note that in all the three naturally aligned scenarios, $\tan\beta$ as given in (\ref{tanb}) `consistently' gives an {\em indefinite} answer 0/0. In what follows, we focus on the simplest realization of the SM alignment, namely, the MS-2HDM based on  the SO(5) group~\cite{Dev:2014yca}. A detailed study of the other two cases will be presented elsewhere. 

%%%%%%%%%%%%%%%%%%%%%%%%%%%%%%%%%%%%%%%%%%%%%%%%%%%%%%%%%%%%%%%%
\section{Maximally Symmetric 2HDM}\label{sec:4}
%%%%%%%%%%%%%%%%%%%%%%%%%%%%%%%%%%%%%%%%%%%%%%%%%%%%%%%%%%%%%%%
From Table~\ref{tab1}, we see that the maximal symmetry  group in the bilinear field space is SO(5), in which case the parameters of the 2HDM potential~\eqref{pot} satisfy the following relations:
\begin{align}
& \mu_1^2 \ = \ \mu_2^2\; , \quad m^2_{12} \ = \ 0\; , \quad \nonumber \\
& \lambda_2 \ = \ \lambda_1\; , \quad 
 \lambda_3  \ = \ 2\lambda_1\; , \quad 
\lambda_4 \ = \  {\rm Re}(\lambda_5) \  = \  \lambda_6 \ = \ \lambda_7\ =\ 0 \; ,
\label{so5}
\end{align} 
Thus, in this case, the 2HDM potential~\eqref{pot} is parametrized by just a {\em single} mass parameter $\mu_1^2=\mu_2^2\equiv \mu^2$ and a {\em single} quartic coupling $\lambda_1=\lambda_2=\lambda_3/2\equiv \lambda$, as in the SM: 
 \begin{align}
   \label{VSO5}
V \ & = \  -\,\mu^2\, \Big(|\Phi_1|^2+|\Phi_2|^2\Big)\: +\: \lambda\,
\Big(|\Phi_1|^2+|\Phi_2|^2\Big)^2 %\nonumber\\
 \  = \ -\: \frac{\mu^2}{2}\, {\bf \Phi}^\dagger\, {\bf \Phi}\ +\ 
\frac{\lambda}{4}\, \big( {\bf \Phi}^\dagger\, {\bf \Phi}\big)^2   \; .
\end{align}
Note   that  the   MS-2HDM   scalar   potential
in~(\ref{VSO5}) is  more minimal than the respective  potential of the
MSSM at the  tree level. Even in the  custodial symmetric limit $g'\to
0$, the latter possesses a smaller symmetry: ${\rm O}(2)\times
{\rm  O}(3)  \subset  {\rm  SO}(5)$,  in  the  5-dimensional  bilinear
$R^I$~space.

Given the isomorphism of the Lie algebras ${\rm SO(5)}  \sim  {\rm Sp}(4)$,\footnote{Here we follow the notation of~\cite{Slansky:1981yr} for denoting the compact, simply connected symplectic group of dimension $n(2n+1)$ as Sp($2n$). In mathematics, this is usually denoted as USp($2n$) or simply as Sp($n$)~\cite{hall}. } the maximal symmetry group  
of the 2HDM in      the      original      {\boldmath     $\Phi$}-field      space
is ${\rm G}^{\bf \Phi}_{\rm 2HDM} = \left[{\rm Sp}(4)/Z_2\right] \times
{\rm SU(2)}_L$~\cite{Pilaftsis:2011ed, Dev:2014yca}\footnote{The quotient factor Z$_2$  is needed to avoid double covering the group ${\rm G}^{\bf \Phi}_{\rm 2HDM}$ in the {\boldmath $\Phi$}-space. Specifically, for each group element $f\in {\rm SU}(2)_L$ and $g\in {\rm Sp}(2n)$, we also have $-f\in {\rm SU}(2)_L$ and $-g\in {\rm Sp}(2n)$, leading to the double-covering equality: $f\otimes g=(-f)\otimes (-g)$.} in the custodial symmetry limit of vanishing  $g'$ and fermion Yukawa couplings. We can generalize this result to deduce that in the custodial symmetry limit, the maximal symmetry group for an $n$ Higgs Doublet Model ($n$HDM) will be
${\rm G}^{\bf \Phi}_{n{\rm HDM}} = \left[{\rm Sp}(2n)/Z_2\right] \times
{\rm SU(2)}_L$.

%%%%%%%%%%%%%%%%%%%%%%%%%%%%%%%%%%%%%%%%%%%%%%%%%%%%%%%%%%%%%%%%%%%%
\subsection{Scalar Spectrum in the MS-2HDM}\label{sec:spec-ms}
%%%%%%%%%%%%%%%%%%%%%%%%%%%%%%%%%%%%%%%%%%%%%%%%%%%%%%%%%%%%%%%%%%%%%%
Using the parameter relations given by \eqref{so5}, we find from~\eqref{mass0}-\eqref{mass3} that in the MS-2HDM, the CP-even Higgs $H$ has mass $M_H^2=2\lambda_2
v^2$, whilst  the remaining four  scalar fields, denoted  hereafter as
$h$,  $a$ and  $h^\pm$, are  massless. This is a consequence of the Goldstone theorem~\cite{goldstone}, since after  electroweak symmetry  breaking, ${\rm  SO}(5) \  \xrightarrow{\langle \Phi_{1,2}\rangle \neq 0}  \ {\rm
  SO}(4)$. Thus, we identify
$H$ as  the SM-like Higgs  boson with the mixing  angle $\alpha=\beta$
[cf.~(\ref{HSM})], i.e. the SM alignment limit can be naturally attributed to the SO(5) symmetry of the theory.  

In the exact SO(5)-symmetric limit, the scalar spectrum of the MS-2HDM
is experimentally unacceptable. This is because the four massless pseudo-Goldstone
particles, viz.~$h$,  $a$ and $h^\pm$,  have sizable couplings to  the 
SM $Z$ and $W^\pm$ bosons, and could induce additional decay channels,
such as~$Z\to  ha$ and $W^\pm  \to h^\pm h$, which  are experimentally
excluded~\cite{PDG}. As we will see in the next subsection,
the   SO(5)  symmetry  may   be  violated
predominantly by  renormalization group (RG) effects due to $g'$  and third-generation Yukawa
couplings, as well as by soft SO(5)-breaking mass parameters, thereby 
lifting the masses of these pseudo-Goldstone particles.

%%%%%%%%%%%%%%%%%%%%%%%%%%%%%%%%%%%%%%%%%%%%%%%%%%%%%%%%%%%%%%%%%%%%%%%%%
\subsection{RG and Soft Breaking Effects \label{sec:RGE}}
\begin{figure*}[t]
\centering
\includegraphics[width=7cm]{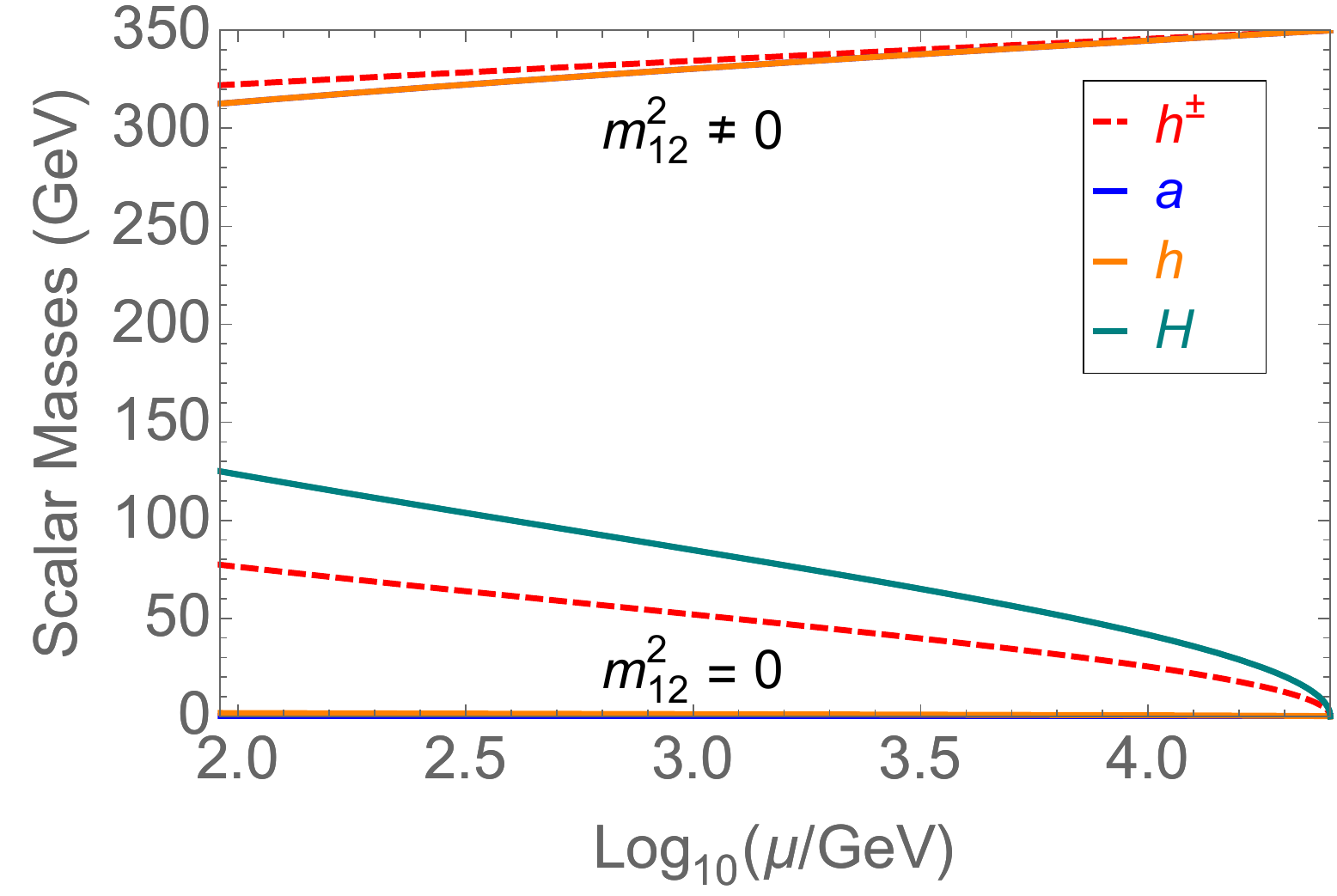}
%\hspace{0.5cm}
%\includegraphics[width=7cm]{fig1b.pdf}
\caption{The Higgs mass spectrum  in the MS-2HDM  without and
  with soft breaking effects  induced by $m^2_{12}$. For $m^2_{12}=0$, the CP-odd scalar $a$ remains massless at tree-level, whereas $h$ and $h^\pm$ receive small masses due to the $g'$ and Yukawa coupling effects. For $m^2_{12}\neq 0$, one obtains a quasi-degenerate heavy Higgs spectrum, cf.~(\ref{mass-so5}). 
%{\em Right--} 
 % The  RG evolution  of the  scalar quartic  couplings under  $g'$ and
  %Yukawa coupling  effects. 
Here we have  chosen $\mu_X=2.5\times 10^{4}$ GeV,
  $\lambda(\mu_X)=0$   and   $\tan\beta   =  50$   for   illustration.
} \label{fig1}
\end{figure*}

%%%%%%%%%%%%%%%%%%%%%%%%%%%%%%%%%%%%%%%%%%%%%%%%%%%%%%%%%%%%%%%%%%%
To calculate the RG and soft-breaking effects in a technically natural manner, we assume that the  SO(5)  symmetry is  realized  at  some  high scale~$\mu_X\gg v$.  
The
physical mass  spectrum at the  electroweak scale is then  obtained by
the RG evolution  of the 2HDM parameters given  by (\ref{pot}).  Using
state-of-the-art  two-loop RG equations given  in~\cite{Dev:2014yca}, 
we  first examine the
deviation of the Higgs spectrum  from the SO(5)-symmetric limit due to
$g'$   and  Yukawa   coupling   effects, in the absence of the soft-breaking term.   This   is  illustrated   in
Figure~\ref{fig1}  for a  typical choice  of parameters  in  a Type-II
realization of the  2HDM. We find that the RG-induced $g'$ effects only lift the charged Higgs-boson mass
$M_{h^\pm}$, while the corresponding Yukawa coupling effects also lift
slightly  the  mass of  the  non-SM CP-even  pseudo-Goldstone
boson~$h$.  However, they still leave the CP-odd scalar $a$ massless,  which can be identified  as a
${\rm  U}(1)_{\rm PQ}$  axion~\cite{Peccei:1977hh}.  

Therefore, $g'$ and Yukawa coupling effects are
{\em  not} sufficient to  yield a  viable Higgs  spectrum at  the weak
scale, starting  from a  SO(5)-invariant boundary condition  at some  high scale
$\mu_X$.   To minimally circumvent  this problem,  we include
soft SO(5)-breaking  effects, by assuming  a non-zero soft-breaking term ${\rm
  Re}(m_{12}^2)$.   
In the  SO(5)-symmetric limit  for the
scalar  quartic couplings,  but  with ${\rm  Re}(m_{12}^2)\neq 0$,  we
obtain the following mass spectrum [cf.~(\ref{mass0})-\eqref{mass3}]:
\begin{eqnarray}
M_H^2 \ = \ 2\lambda_2 v^2\; , \qquad M_h^2 \ = \ M_a^2 \ = \ M^2_{h^\pm} \ = \
\frac{{\rm Re}(m^2_{12})}{s_\beta c_\beta} \; ,
\label{mass-so5}
\end{eqnarray}
as  well as  an equality  between  the CP-even  and CP-odd  mixing
angles: $\alpha = \beta$, thus predicting an {\it exact} alignment for
the  SM-like  Higgs  boson  $H$, simultaneously with  
an experimentally allowed heavy Higgs spectra (see Figure~\ref{fig1} for $m^2_{12}\neq 0$ case). Note that in the alignment limit, the heavy Higgs sector is exactly degenerate [cf.~(\ref{mass-so5})] at the SO(5) symmetry-breaking scale, and at the low-energy scale, this degeneracy is mildly broken by the RG effects. Thus, we obtain a quasi-degenerate heavy Higgs spectrum, which is a unique prediction of the MS-2HDM, valid even in the non-decoupling limit, and can be used to distinguish this model from other 2HDM scenarios. 

%%%%%%%%%%%%%%%%%%%%%%%%%%%%%%%%%%%%%%%%%%%%%%%%%%%%%%%%%%%%
\subsection{Misalignment Predictions \label{sec:MP}}
\begin{figure}[t!]
\centering
\includegraphics[width=7cm]{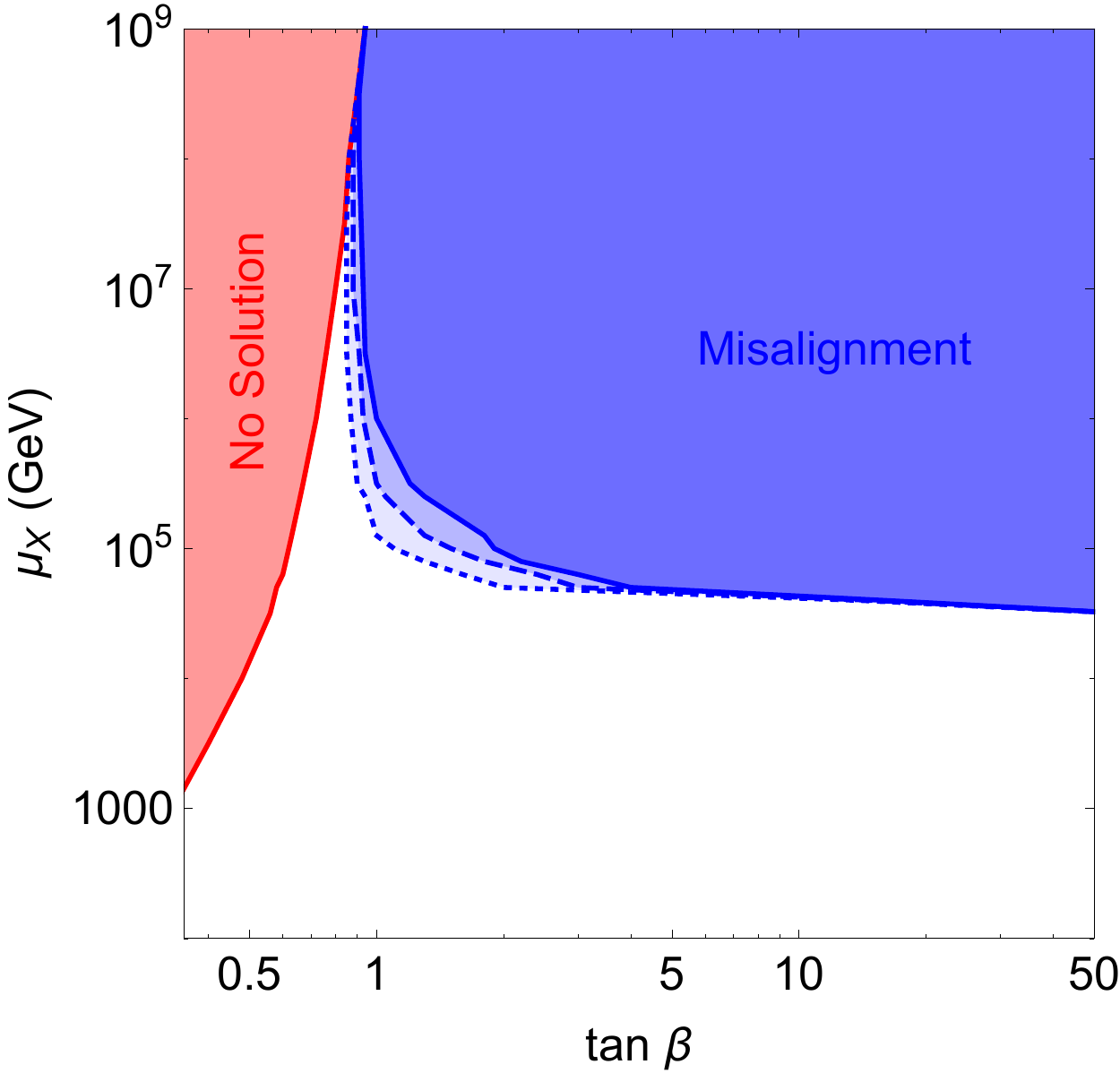}
\caption{The 
$1\sigma$ (dotted), $2\sigma$ (dashed) and $3\sigma$ (solid) exclusion contours (blue shaded region) 
from the alignment constraints in MS-2HDM. The red shaded region is theoretically excluded, as there is no solution to the RG equations up to two-loop order in this region.%, since the RGEs given in Appendix~\ref{app:RGE} do not have a solution 
%in this region. 
} \label{fig2} 
\end{figure}

%%%%%%%%%%%%%%%%%%%%%%%%%%%%%%%%%%%%%%%%%%%%%%%%%%%%%%%%%%%%
As discussed in Section~\ref{sec:RGE}, there will be some deviation
from  the  alignment  limit  in  the low-energy  Higgs  spectrum of the MS-2HDM due to RG and soft-breaking effects.   By
requiring that the  mass and couplings of the  SM-like Higgs boson $H$  
are consistent with the latest Higgs  data from the 
LHC~\cite{coup}, we derive predictions for the remaining
scalar spectrum and compare them with the existing (in)direct limits on the heavy
Higgs sector. For the  SM-like Higgs boson mass,  we use  the $3\sigma$
allowed range from the recent CMS and ATLAS
Higgs    mass    measurements~\cite{coup, Aad:2014aba}: 
$M_H \in  \big[124.1,~ 126.6\big]~{\rm GeV}$. 
For the Higgs  couplings to the SM vector bosons  and fermions, we use
the constraints in  the $(\tan\beta,~\beta-\alpha)$ plane derived from
a       recent       global       fit      for       the       Type-II
2HDM~\cite{Eberhardt:2013uba}. For   a  given  set  of   SO(5)  boundary  conditions
$\big\{\mu_X,\tan\beta(\mu_X),\lambda(\mu_X)\big\}$, we  thus require that the
RG-evolved 2HDM  parameters at the  weak scale must satisfy  the above
constraints  on the  lightest  CP-even Higgs  boson sector.   This
requirement of {\it alignment} with the SM Higgs sector puts stringent
constraints   on   the   MS-2HDM   parameter  space,   as   shown   in
Figure~\ref{fig2} by the blue shaded
region.    
In the red  shaded    region, there is no viable solution to the RG equations. 
We  ensure  that  the  remaining  allowed (white) region  satisfies  the
necessary theoretical constraints,  i.e.~positivity and vacuum  stability of the
Higgs     potential,    and     perturbativity     of    the     Higgs
self-couplings~\cite{review}.   From Figure~\ref{fig2},  we  find that
there  exists  an upper  limit  of  $\mu_X\lesssim  10^9$ GeV  on  the
SO(5)-breaking  scale   of  the   2HDM  potential,  beyond   which  an
ultraviolet completion  of the theory must be  invoked.  Moreover, for
$10^5~{\rm GeV}\lesssim \mu_X \lesssim  10^9~{\rm GeV}$, only a narrow
range of $\tan\beta$ values are allowed.

\begin{figure}[t!]
\centering
\includegraphics[width=7cm]{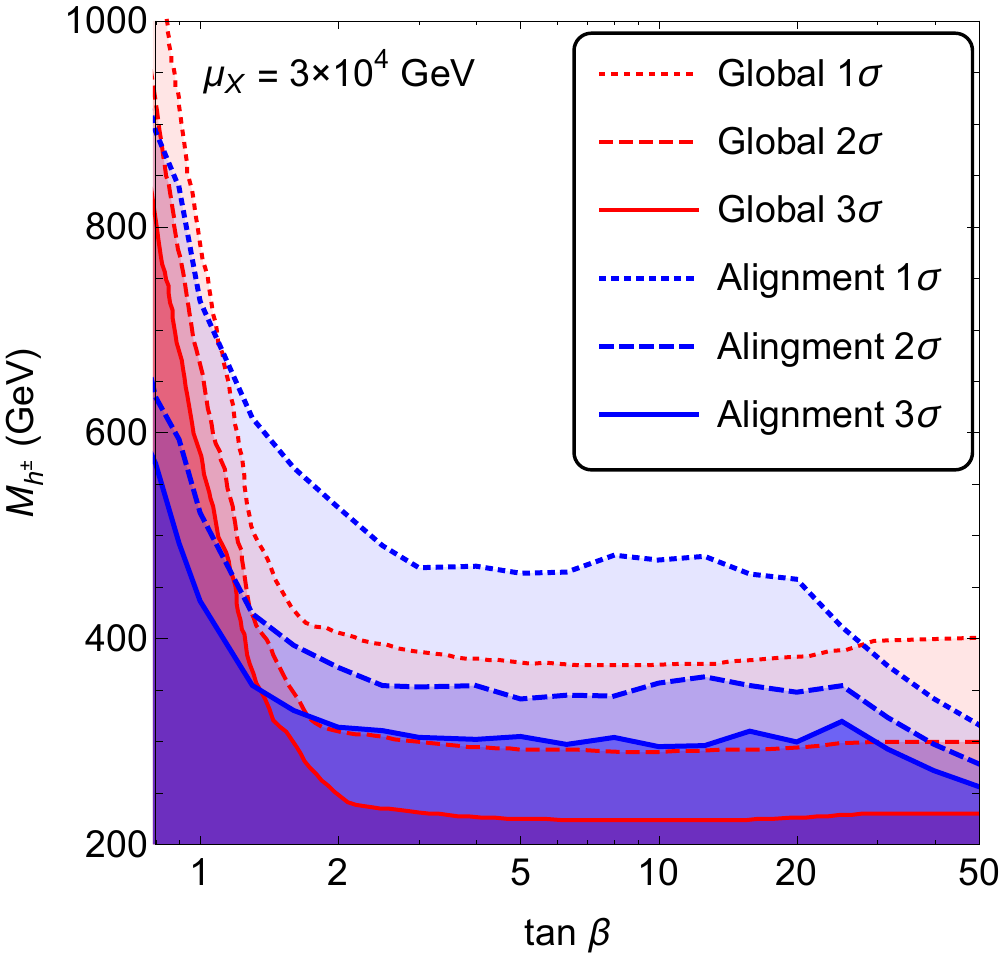}
\includegraphics[width=7cm]{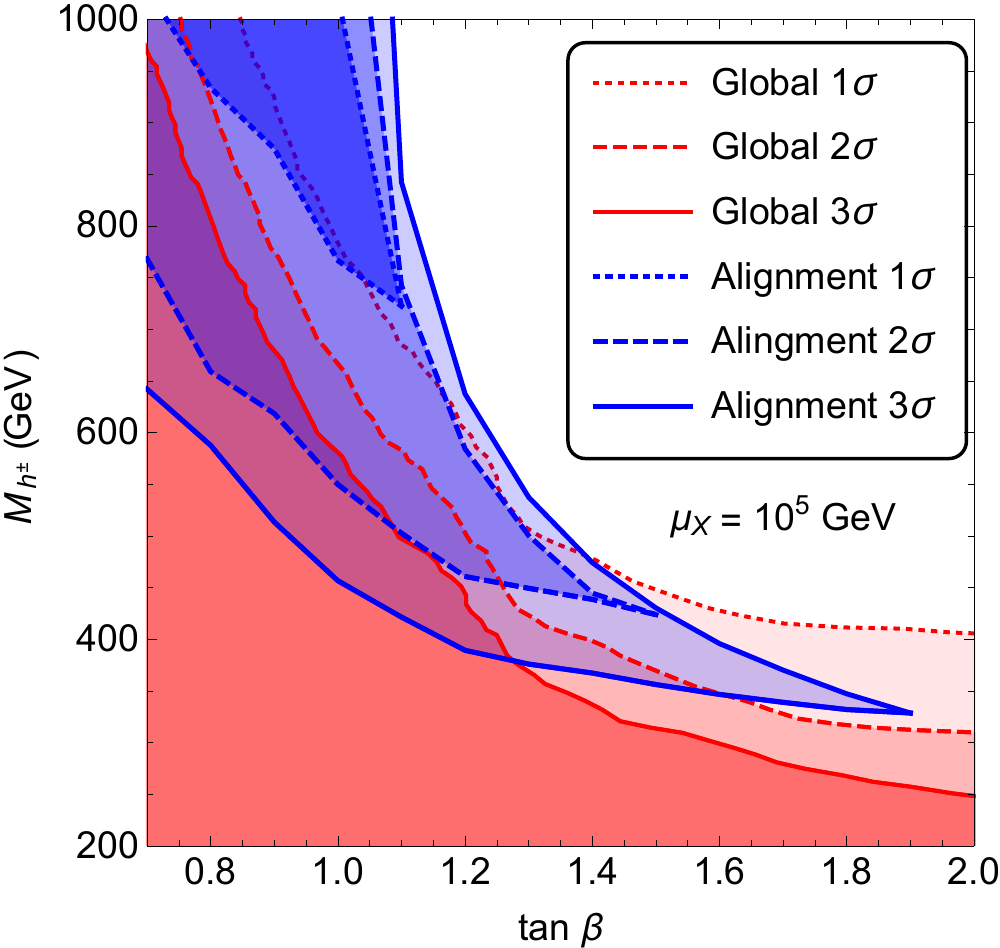}
\caption{{\em Left:} The $1\sigma$ (dotted), $2\sigma$ (dashed) and $3\sigma$ (solid) {\em lower} limits 
on the charged Higgs mass obtained from the alignment condition (blue lines) 
in the MS-2HDM with $\mu_X=3\times 10^{4}$ GeV. {\em Right:}  The $1\sigma$ (dotted), $2\sigma$ (dashed) and $3\sigma$ (solid) {\em allowed} regions from the alignment condition (blue lines) for  $\mu_X=10^{5}$ GeV. 
For comparison, the corresponding lower limits from a global fit 
are also shown (red lines). } \label{mhp}
\end{figure}

For  the  allowed   parameter  space  of  our  MS-2HDM   as  shown  in
Figure~\ref{fig2},  we obtain concrete  predictions for  the remaining
Higgs  spectrum.  In  particular,  the alignment  condition imposes  a
{\it lower} bound on the  soft breaking parameter Re$(m^2_{12})$, and hence,
on the heavy  Higgs spectrum. The  comparison of  the
existing  global fit limit on  the  charged  Higgs-boson mass  as  a function  of
$\tan\beta$~\cite{Eberhardt:2013uba} with our predicted limits from the alignment condition in the MS-2HDM for
a typical value of the boundary scale $\mu_X=3\times 10^{4}$ GeV is shown in Figure~\ref{mhp} (left panel).  It
is  clear that  the alignment  limits are  stronger than  the global fit 
limits, except  in the very small  and very large $\tan\beta$ regimes. For 
$\tan\beta\lesssim 1$ region, the indirect limit obtained  from 
the $Z\to b\bar{b}$  precision observable becomes
the strictest~\cite{Deschamps:2009rh, Eberhardt:2013uba}. Similarly, for the 
large $\tan\beta\gtrsim 30$ case, the alignment limit can be easily obtained 
[cf.~(\ref{chat})] 
without requiring a large soft-breaking parameter $m_{12}^2$, and therefore, 
the lower limit on the charged Higgs mass derived from the misalignment condition 
becomes somewhat weaker in this regime. 

From Figure~\ref{fig2}, it should be noted that for $\mu_X\gtrsim
10^5$~GeV, phenomenologically acceptable alignment is not possible in the MS-2HDM 
for large $\tan\beta$ {\it and} large $m^2_{12}$.  Therefore, we also get an   {\it  upper}  bound  on  the  charged
Higgs-boson mass  $M_{h^\pm}$ from the misalignment condition, depending on $\tan\beta$. This is illustrated in Figure~\ref{mhp} (right panel)  
for $\mu_X=10^5$ GeV.

Similar alignment constraints are obtained  for the heavy neutral pseudo-Goldstone
bosons $h$ and  $a$, which are predicted to be quasi-degenerate with the charged Higgs boson $h^\pm$ in the MS-2HDM [cf.~(\ref{mass-so5})]. The current experimental lower limits on the heavy neutral Higgs sector~\cite{PDG} are much weaker than the alignment constraints in this case. Thus, the MS-2HDM scenario provides a natural reason for the absence of a heavy Higgs signal below the top-quark threshold, and this has important consequences for the heavy Higgs searches in the run-II phase of the LHC, as discussed in the following section.

\section{Collider Signatures in the Alignment Limit} \label{sec:5}

In the alignment limit, the  couplings of the lightest CP-even Higgs
boson are exactly  similar to the SM Higgs  couplings, while the heavy
CP-even    Higgs    boson  is gaugephobic [cf.~\eqref{coup1}].   Therefore,  two   of  the  relevant  Higgs
production mechanisms at the LHC,  namely, the vector boson fusion and
Higgsstrahlung processes are suppressed  for the heavy neutral Higgs sector. 
As a consequence, the only relevant production channels to probe the neutral Higgs
sector  of the  MS-2HDM  are the  gluon-gluon  fusion and  $t\bar{t}h$ 
($b\bar{b}h$) associated  production mechanisms at low (high) $\tan\beta$. 
For  the  charged Higgs sector of the MS-2HDM, the  dominant production 
mode is the associated production process: $gg\to \bar{t}bh^++t\bar{b}h^-$, irrespective of 
$\tan\beta$.

Similarly, for the decay modes of the heavy neutral Higgs bosons in the MS-2HDM, 
the $t\bar{t}$ ($b\bar{b}$) channel is the  dominant one for low (high) $\tan\beta$ values, 
whereas for the charged Higgs boson $h^{+(-)}$, the $t\bar{b}(\bar{t}b)$ mode is the 
dominant one for any $\tan\beta$. Thus, the heavy Higgs sector of the MS-2HDM can be effectively probed at the LHC through the final states involving third-generation quarks. 

%%%%%%%%%%%%%%%%%%%%%%%%%%%%%%%%
\subsection{Charged Higgs Signal} \label{sec5.2}
\begin{figure}[t]
\centering
\includegraphics[width=7cm]{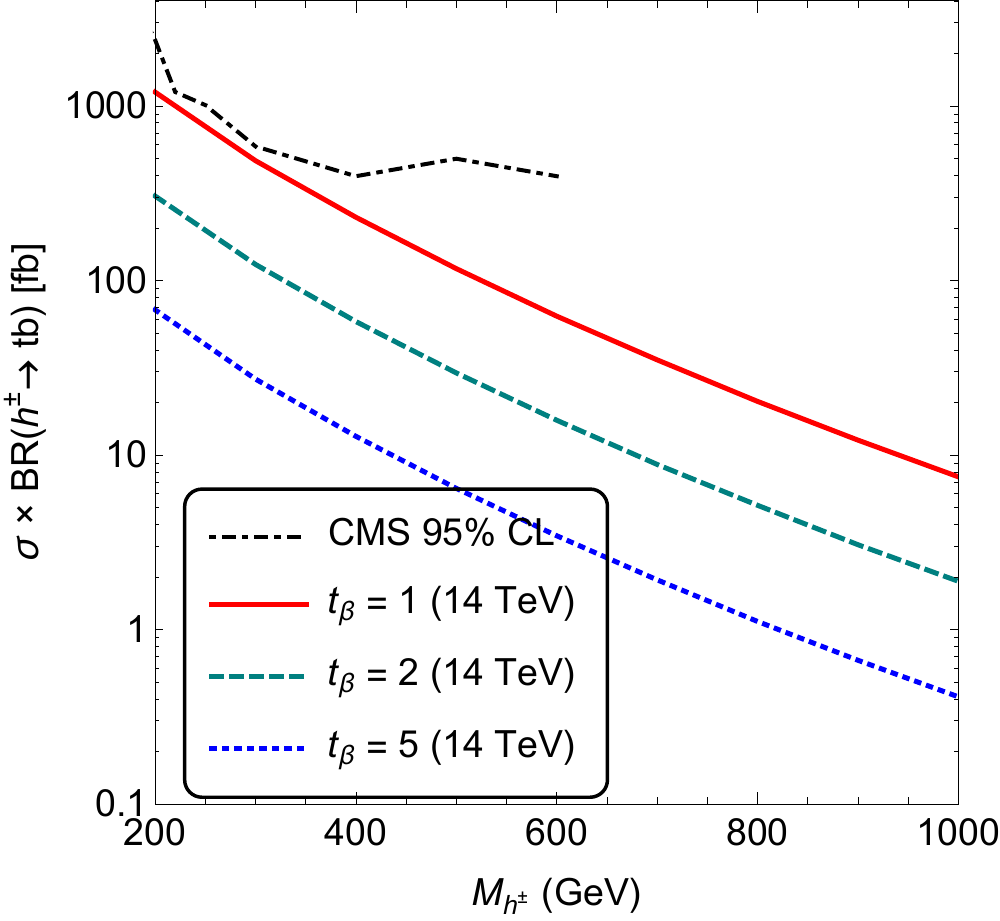}
\caption{Predictions for the cross section of the process (\ref{ttbb}) in the Type-II MS-2HDM 
at $\sqrt s=14$ TeV LHC for various values of $\tan\beta$. For comparison, we have also shown the current 95\% CL CMS upper limit from the $\sqrt s=8$ TeV data~\cite{tb}.}    
\label{ttbb-cross}
\end{figure}

%%%%%%%%%%%%%%%%%%%%%%%%%%%%%%%%
The most promising channel at the LHC for the charged Higgs boson in the MS-2HDM is 
\begin{equation}
gg\ \to\ \bar{t} b h^+ + t\bar{b}h^- \ \to\ t\bar{t} b \bar{b}\;.
\label{ttbb}
\end{equation}
Experimentally, this is a challenging mode due to 
large  QCD backgrounds and the non-trivial event topology, involving at least four 
$b$-jets~\cite{hwg}. Nevertheless, a recent CMS study~\cite{tb} has presented for the first time a realistic analysis of this process, in the leptonic decay mode of the $W$'s coming from top decays: 
\begin{equation}
gg\ \to\ h^\pm tb \ \to \ (\ell \nu_\ell bb)(\ell'\nu_{\ell'}b)b 
\label{ttbb-ll}
\end{equation}
($\ell,\ell'$ beings electrons or muons). Using the $\sqrt s=8$ TeV LHC data, they have derived 95\% CL upper limits on the production cross section $\sigma(gg\to h^\pm tb)$ times the branching ratio BR($h^\pm\to tb$) as a function of the charged Higgs mass, as shown in Figure~\ref{ttbb-cross}. In the same Figure, we show the corresponding predictions at $\sqrt s=14$ TeV LHC in the Type-II MS-2HDM for some representative values of $\tan\beta$. The cross section predictions were obtained at leading order (LO) by implementing the 2HDM in {\tt MadGraph5\_aMC@NLO}~\cite{mg5} and using the {\tt NNPDF2.3} PDF sets~\cite{nnpdf}.  A comparison of these cross sections with the CMS limit suggests that the run-II phase of the LHC might be able to probe the low $\tan\beta$ region of the MS-2HDM parameter space using the process (\ref{ttbb}).  Note that the production
cross  section  $\sigma(gg\to  \bar{t}  b  h^+)$ decreases  rapidly  with
increasing $\tan\beta$ due to the Yukawa coupling suppression, 
even though  BR($h^+\to
\bar{t}b$)  remains close to  100\%. Therefore, this channel is only effective for low $\tan\beta$ values. 

In order to make a rough estimate of the $\sqrt s=14$ TeV LHC sensitivity to the charged Higgs signal (\ref{ttbb}) in the MS-2HDM, we perform a parton level simulation of the signal and background events using {\tt MadGraph5}~\cite{mg5}. For the event reconstruction, we use 
some basic selection  cuts on  the transverse
momentum, pseudo-rapidity and dilepton invariant mass, following the CMS analysis~\cite{tb}:
\begin{eqnarray}
&& p_T^\ell \ > \ 20~{\rm GeV}, \quad |\eta^\ell| < 2.5,  \quad p_T^j \ > \ 30~{\rm GeV}, \quad |\eta^j| < 2.4, \quad \slashed{E}_T > 40~{\rm GeV}\nonumber \\
&& \Delta R^{\ell\ell} > 0.4, \quad \Delta R^{\ell j} > 0.4, \quad M_{\ell\ell} > 12~{\rm GeV}, \quad |M_{\ell\ell}-M_Z| > 10~{\rm GeV}.
\label{cut}
\end{eqnarray}
Jets are reconstructed using the anti-$k_T$ clustering algorithm~\cite{anti-kT} with a distance parameter of 0.5. Since four $b$-jets are expected in the final state, at least two $b$-tagged jets are required in the signal events, and we assume the $b$-tagging efficiency for each of them to be  70\%.

\begin{figure}[t]
\centering
\includegraphics[width=7cm]{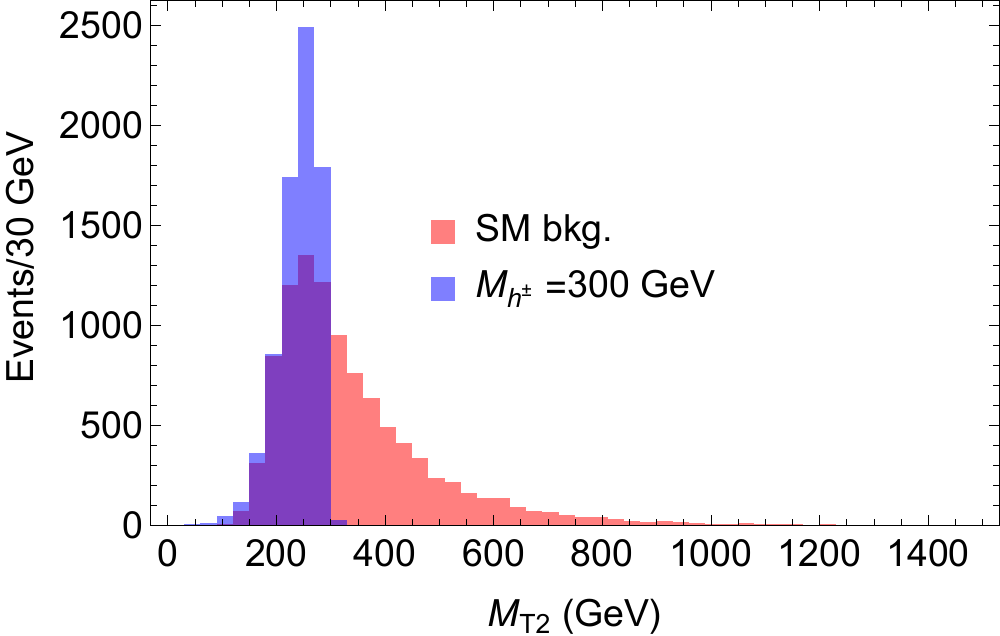}
\caption{An illustration of the charged Higgs boson mass reconstruction using the $M_{T2}$ variable. The irreducible SM background distribution is also shown for comparison.}    
\label{ttbb-dist}
\end{figure}

The inclusive SM  cross section for $pp\to t\bar{t}b\bar{b}+X$  is $\sim 18$ pb at NLO, with roughly 30\% uncertainty due to higher order QCD  corrections~\cite{pittau}. Most of the QCD background for the $4b+2\ell+\slashed{E}_T$ final state given by (\ref{ttbb-ll}) can be reduced significantly by reconstructing at least one top-quark. The remaining irreducible background due to SM $t\bar{t}b\bar{b}$ production can be suppressed with respect to the signal by reconstructing the charged Higgs boson mass, once a valid signal region is defined, e.g. in terms of an observed excess of events at the LHC in future.  For the semi-leptonic decay mode of top-quarks as in (\ref{ttbb-ll}), one cannot directly use an invariant mass observable to infer $M_{h^\pm}$, as both the neutrinos in the final state give rise to missing momentum. A useful quantity in this case is the $M_{T2}$ variable~\cite{mt2}, 
defined as 
\begin{eqnarray}
M_{T2} \ = \ \underset{\left\{ \slashed{\mathbf p}_{T_{\rm a}}+\slashed{\mathbf p}_{T_{\rm b}}=\slashed{\mathbf p}_T\right\}}{\rm min}\Big[{\rm max}\left\{m_{T_{\rm a}},m_{T_{\rm b}}\right\}\Big] \;,
\label{mt2}
\end{eqnarray}
 where $\{{\rm a}\}, \{\rm b\}$ stand for the two sets of particles in the final state, each containing a neutrino with part of the missing transverse momentum ($\slashed{\mathbf p}_{T_{\rm {a,b}}}$). Minimization over all possible sums of these two momenta gives the observed missing transverse momentum $\slashed{\mathbf p}_T$, whose magnitude is the same as $\slashed{E}_T$ in our specific case. In (\ref{mt2}), $m_{T_{i}}$ (with $i=$a,b) is the usual transverse mass variable for the system $\{i\}$, defined as 
\begin{eqnarray}
m_{T_{i}}^2 \ = \ \left(\sum_{\rm visible} E_{T_i}+\slashed{E}_{T_i} \right)^2- \left(\sum_{\rm visible} {\mathbf p}_{T_i}+\slashed{\mathbf p}_{T_i} \right)^2 \; .
\end{eqnarray}
For the correct combination of the final state particles in (\ref{ttbb-ll}), i.e. for $\{{\rm a}\}=(\ell \nu_\ell bb)$ and $\{{\rm b}\}=(\ell' \nu_{\ell'}bb)$ in (\ref{mt2}), the maximum value of 
$M_{T2}$ represents the charged Higgs boson mass, with the $M_{T2}$ distribution smoothly dropping to zero at this point. This is illustrated in Figure~\ref{ttbb-dist} for a typical choice of $M_{h^\pm}=300$ GeV. For comparison, we also show the $M_{T2}$ distribution for the SM background, which obviously does not have a sharp endpoint. Thus, for a given hypothesized signal region defined in terms of an excess due to $M_{h^\pm}$, we may impose an additional cut on $M_{T2}\leq M_{h^\pm}$ to enhance the signal (\ref{ttbb-ll}) over the irreducible SM background. 

\begin{figure}[t]
\centering
\includegraphics[width=7cm]{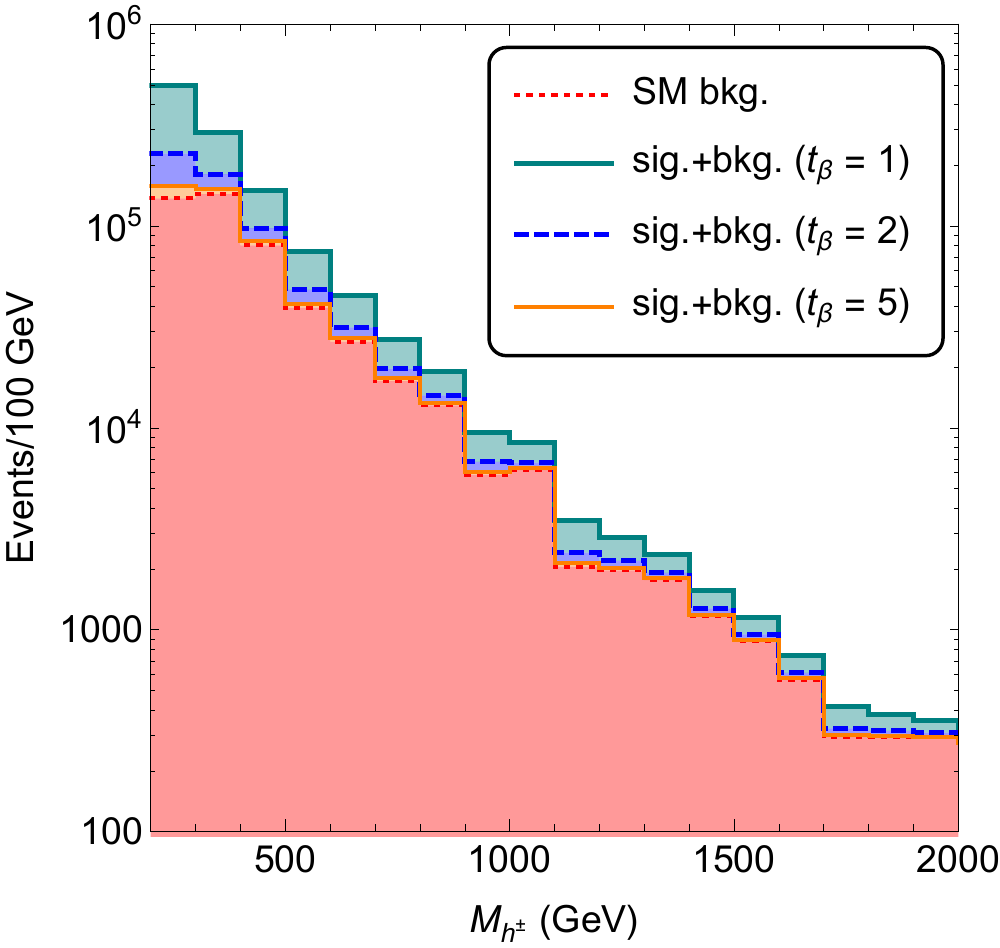}
\caption{Predicted number of events for the dominant charged Higgs signal in the MS-2HDM at $\sqrt
  s=14$ TeV LHC with $300~{\rm fb}^{-1}$ integrated luminosity. The irreducible SM background (red shaded) is controlled by assuming an efficient mass reconstruction technique~\cite{Dev:2014yca}. } 
\label{2tb}
\end{figure}

Assuming that the charged Higgs boson mass can be reconstructed efficiently, we present an estimate of the signal to background ratio for the charged Higgs signal given by (\ref{ttbb}) at $\sqrt s=14$ TeV LHC with 300 fb$^{-1}$ for some typical values of $\tan\beta$ in Figure~\ref{2tb}. Since the mass of the charged Higgs boson is a priori unknown, we vary the charged Higgs mass, and for each value of $M_{h^\pm}$, we assume that it can be reconstructed around its actual value within 30 GeV uncertainty.

\subsection{Heavy Neutral Higgs Signal} \label{sec5.3}

So far there have been no direct searches for heavy neutral Higgs bosons involving $t\bar{t}$ and/or $b\bar{b}$ final states, mainly due to the challenges associated with uncertainties in the jet energy scales and the combinatorics arising from complicated multiparticle final states in a busy QCD environment. Nevertheless, these channels become pronounced in the MS-2HDM scenario, and hence, we have made a preliminary attempt to study them in~\cite{Dev:2014yca}. In particular, we focus on the search channel  
\begin{equation}
gg\ \to\  t\bar{t}h\  \to\ t\bar{t}t\bar{t}\; .
\label{4tp}
\end{equation}
Such four top final states have been proposed before in the context of
other  exotic  searches   at  the  LHC (see e.g.~\cite{4topother}).  However,  their relevance for heavy Higgs searches have not been explored so far. We note here that the existing 95\% CL experimental upper limit on the four top production cross section is 59 fb from ATLAS~\cite{atlas-4t} and 32 fb from CMS~\cite{cms-4t}, whereas the SM prediction for the inclusive cross section of the process $pp\to t\bar{t}t\bar{t}+X$ is about 10-15 fb~\cite{Bevilacqua:2012em}.

\begin{figure}[t!]
\centering
\includegraphics[width=7cm]{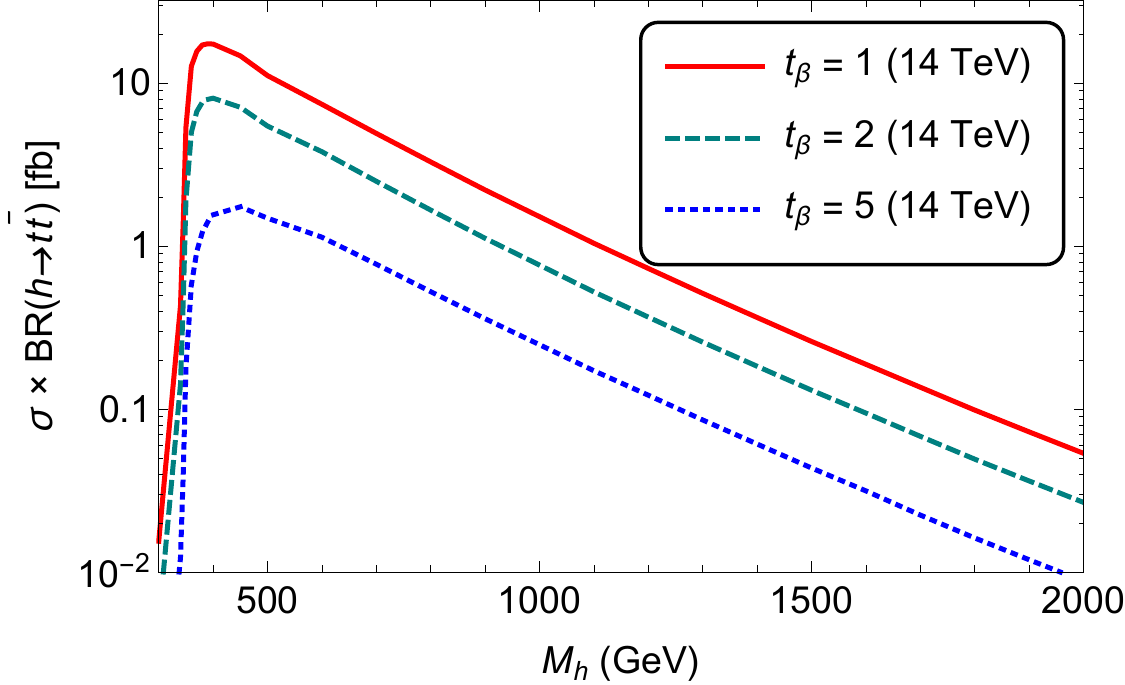}
\caption{Predictions for the cross section of the process (\ref{4tp}) in the Type-II MS-2HDM at $\sqrt s=14$ TeV LHC for various values of $\tan\beta$. } \label{4tx}
\end{figure}

To get a rough estimate of  the signal to background ratio for our 
four-top signal~\eqref{4tp},  we perform a  parton-level simulation of  the signal
and   background  events   at  LO  in   QCD   using  {\tt
  MadGraph5\_aMC@NLO}~\cite{mg5}   with   {\tt NNPDF2.3}  PDF sets~\cite{nnpdf}.  
For the  inclusive  SM  cross  section  for  the
four-top final  state at $\sqrt s=14$ TeV LHC, we obtain 11.85 fb, whereas  our proposed four-top 
signal  cross sections are
found   to  be   comparable  or   smaller  depending   on   $M_h$  and
$\tan\beta$, as shown in Figure~\ref{4tx}. However, since  we expect  one of  the  $t\bar{t}$ pairs
coming from  an on-shell  $h$ decay to  have an invariant  mass around
$M_h$,  we can  use this information to significantly boost the signal over the irreducible SM background. Note that all the predicted cross sections shown in Figure~\ref{4tx} are well below 
the current  experimental  upper bound~\cite{cms-4t}. 

Depending on the $W$ decay mode from $t\to Wb$, there are 35 final states for four top decays. According to a recent ATLAS analysis~\cite{thesis}, the experimentally favored channel is the semi-leptonic/hadronic final state with two same-sign isolated leptons. Although the branching fraction for this topology (4.19\%) is smaller than most of the other channels, the presence of two same-sign leptons in the final state allows us to reduce the large QCD background substantially, including that due to the SM production of $t\bar{t}b\bar{b}+$jets~\cite{thesis}. Therefore, we will only consider the following decay chain in our preliminary analysis: 
\begin{eqnarray}
gg\ \to\  t\bar{t}h\  \to\ (t\bar{t})(t\bar{t}) \ \to \  
\Big( (\ell^\pm \nu_{\ell}b)(jjb)\Big)\Big((\ell'^\pm \nu_{\ell'}b)(jjb)\Big)\; .
\label{4t-ll}
\end{eqnarray}
For event reconstruction, we will use the same selection cuts as in (\ref{cut}), and in addition, following~\cite{thesis}, 
we require the scalar sum of the $p_T$ of all leptons and jets (defined as $H_T$) to exceed 350  GeV.    

\begin{figure}[t]
\centering
\includegraphics[width=7cm]{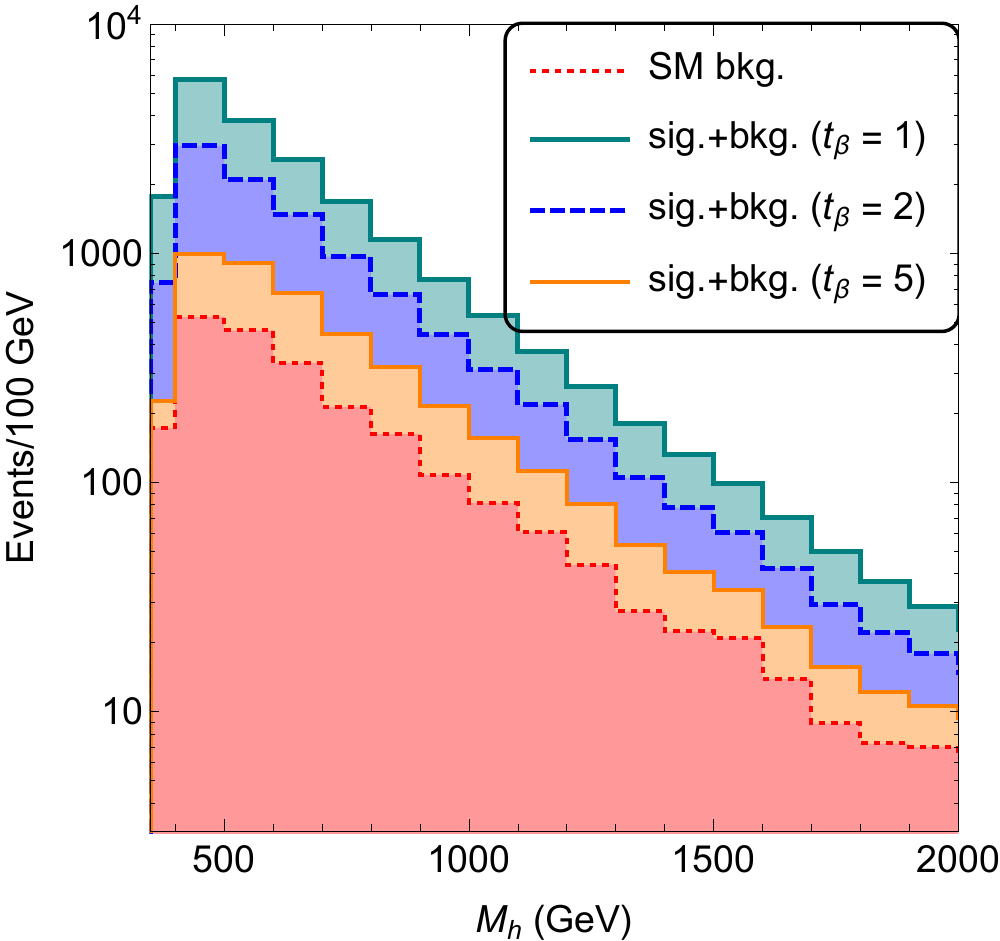}
\caption{Predicted number of  events for the $t\bar{t}t\bar{t}$ signal
  from  the neutral pseudo-Goldstone  boson in  the MS-2HDM  at $\sqrt
  s=14$ TeV  LHC with  $300~{\rm fb}^{-1}$ integrated  luminosity. } \label{4t}
\end{figure}

As in the charged Higgs boson case (cf.~Figure~\ref{ttbb-dist}), the heavy Higgs mass can be reconstructed from the signal given by (\ref{4t-ll}) using the $M_{T2}$ endpoint technique, and therefore, an additional selection cut on $M_{T2}\leq M_h$ can be used to enhance the signal over the irreducible background. 
Our simulation results  for the predicted number of signal and background events for the process (\ref{4t-ll}) at $\sqrt s=14$ TeV LHC with 300 fb$^{-1}$ luminosity are  shown in Figure~\ref{4t}. The signal events are shown for  three representative
values of $\tan\beta$. Here we vary the a priori unknown heavy Higgs mass, and for each value of $M_{h}$, we assume that it can be reconstructed around its actual value within 30 GeV uncertainty. From this preliminary
analysis, we  find that the
$t\bar{t}t\bar{t}$ channel provides the most promising collider signal
to  probe the  heavy Higgs  sector in  the MS-2HDM  at low  values of
$\tan\beta \lesssim  5$.

The above analysis  is also applicable for the  CP-odd Higgs boson
$a$,  which  has  similar  production  cross  sections  and  $t\bar{t}$
branching fractions as the CP-even Higgs $h$. However, 
the $t\bar{t}h(a)$ production cross section  as well as the $h(a)\to t\bar t$ branching
ratio decreases with  increasing $\tan\beta$. This is due  to the fact
that   the   $ht\bar{t}$   coupling   in  the   alignment   limit   is
$\cos\alpha/\sin\beta\sim \cot\beta$, which is same as the $at\bar{t}$
coupling. Thus,  the high $\tan\beta$ region of
the MS-2HDM  cannot be searched  via the $t\bar{t}t\bar{t}$ channel  proposed above,
and one  needs to consider  the channels involving  down-sector Yukawa
couplings, e.g. $b\bar{b}b\bar{b}$ and $b\bar{b}\tau^+\tau^-$~\cite{hwg}. It is also worth commenting here that the simpler process  $pp\to h/a \to t\bar{t}~(b\bar{b})$ at low (high) $\tan\beta$ suffers from a huge SM $t\bar{t}$ ($b\bar{b}$) QCD background, even after imposing an $M_{t\bar{t}~(b\bar{b})}$ cut. Some parton-level studies of this signal in the context of MSSM have been performed in~\cite{Djouadi:2013vqa}.

We should clarify that the results obtained in this section are valid only at the parton level. In a realistic detector environment, the sharp features of the signal [see e.g.,~Figure~\ref{ttbb-dist}] used to derive the sensitivity reach in  Figures~\ref{2tb} and \ref{4t}  may not survive, and therefore, the signal-to-background ratio might get somewhat reduced than that shown here. A detailed detector-level  analysis  of  these  signals,  including realistic  top reconstruction  efficiencies and smearing effects, is currently being  pursued in  a separate dedicated study.

\section{Conclusions}\label{sec:6}

We provide a symmetry justification of the so-called SM alignment limit, independently of
the  heavy  Higgs  spectrum  and  the  value  of  $\tan\beta$ in the  2HDM.  
We show that in the 2HDMs where both Higgs doublets acquire VEVs, there exist {\em only} three different symmetry realizations, which could lead to the SM alignment by satisfying the natural alignment condition~(\ref{alcond}) for {\em any} value of $\tan\beta$. In the context of the Maximally Symmetric 2HDM  based  on the  SO(5)  group,  we demonstrate how small
deviations from this alignment limit are naturally induced by RG 
effects  due  to  the   hypercharge  gauge  coupling  $g'$  and  third
generation Yukawa  couplings, which explicitly break  the custodial symmetry
of  the theory.   In  addition, a  non-zero  soft SO(5)-breaking  mass
parameter is required to yield a viable Higgs spectrum consistent with
the  existing experimental constraints.   Using the  current Higgs
signal strength  data from the  LHC, which disfavor  large deviations
from the alignment limit, we  derive important constraints on the 2HDM
parameter space.  In particular, we  predict lower limits on the mass scale of the heavy
Higgs spectrum,  which prevail the present  global fit limits in a  wide range of
parameter space.   Depending on the  energy scale where the  maximal symmetry
could be  realized in  nature, we  also obtain an  upper limit  on the
heavy Higgs masses in certain  cases, which could be probed
during  the  run-II phase  of  the LHC.   In addition,  we  have studied the collider signatures of the heavy Higgs sector in the alignment limit  beyond the top-quark threshold. We find that the final states involving third-generation quarks can
become a valuable observational tool to directly probe the heavy Higgs
sector of the 2HDM in the alignment limit for low values of $\tan\beta$. Finally, we emphasize the importance of {\em both} charged and neutral heavy Higgs searches in order to unravel the {\em doublet} nature of the heavy Higgs sector.

%\vfill\eject

\section*{Acknowledgments}
This work was  supported   by  the   Lancaster-Manchester-Sheffield
Consortium for Fundamental Physics under STFC grant ST/L000520/1. P.S.B.D. would like to acknowledge the local hospitality provided by the CFTP and IST, Lisbon where part of this article was written.

\end{document}